# Illuminating the electronic properties of WS$_2$ polytypism with electron microscopy

Sabrya E. van Heijst[1], Masaki Mukai[2], Eiji Okunishi[2], Hiroki Hashiguchi[2], Laurien I. Roest[1,4], Louis Maduro[1], Juan Rojo[4,5], and Sonia Conesa-Boj[1,*].

[1] Kavli Institute of Nanoscience, Delft University of Technology, 2628CJ, Delft, The Netherlands.

[2] EMBU JEOL Ltd, Tokyo, Japan.

[4] Nikhef Theory Group, Science Park 105, 1098XG Amsterdam, The Netherlands.

[5] Department of Physics and Astronomy, VU Amsterdam, 1081 HV Amsterdam, The Netherlands.

* e-mail: s.conesaboj@tudelft.nl

**ABSTRACT**

Tailoring the specific stacking sequences (polytypes) of layered materials represents a powerful strategy to identify and design novel physical properties. While nanostructures built upon transition-metal dichalcogenides (TMDs) with either the 2H or 3R crystalline phases have been routinely studied, our knowledge of those based on mixed 2H/3R polytypes is far more limited. Here we report on the characterization of mixed 2H/3R free-standing $WS_2$ nanostructures displaying a flower-like configuration by means of advanced transmission electron microscopy. We correlate their rich variety of shape-morphology combinations with relevant local electronic properties such as their edge, surface, and bulk plasmons. Electron energy-loss spectroscopy combined with machine learning reveals that the 2H/3R polytype displays an indirect band gap with $E_{BG} = 1.6^{+0.3}_{-0.2}$ eV. Further, we identify the presence of energy-gain peaks in the EEL spectra characterized by a gain-to-loss ratio $I_G/I_L > 1$. Such property could be exploited to develop novel cooling strategies for atomically thin TMD nanostructures and devices built upon them. Our results represent a stepping stone towards an improved understanding of TMDs nanostructures based on mixed crystalline phases.

# INTRODUCTION

Many of the remarkable electronic and optical properties exhibited by transition-metal dichalcogenide (TMD) materials can be traced back to the underlying periodic arrangements of their layers, the so-called stacking sequences[1,2]. Therefore, identifying and controlling these stacking sequences provides a powerful handle in the quest to design novel TMD-based nanostructures with tailored functionalities.

The most common stacking sequences present in TMDs are the 2H and the 3R polytypes[3,4,5]. The 2H phase belongs to the space group P63/mmc and has a unit cell composed of a bilayer following the AA' stacking sequence, a configuration characterized by an inversion symmetry[6]. The 3R phase belongs instead to the space group R3m and is defined by a bilayer with an AB stacking sequence, which as opposed to the 2H phase does not exhibit inversion symmetry[7]. Both crystalline phases are known to display a semiconducting behavior[8]. While several studies of the structural, optical, and electronic properties of TMDs based on either the 2H or the 3R phases have been reported[9,10,11], much less is known about how these properties are modified in the presence of a mixed 2H/3R polytypism[12]. Unraveling the implications of such polytypism in TMDs would open new avenues in applications from nanoplasmonics and nanoelectronics to catalysis[13,14,15].

In this work, we report on the characterization of $WS_2$ flower-like nanostructures ("nanoflowers") composed by randomly oriented flakes (the "petals") arising from a common point (the "stem"). These $WS_2$ nanoflowers display a rich variety of shape-morphology configurations, such as lying petals and edge-exposed standing petals. Together with their polytypism, this unique feature turns these nanostructures into an ideal laboratory for the study of the modifications of local electronic properties in $WS_2$.

First of all, aberration-corrected scanning transmission electron microscopy (AC-STEM)[16] is used to reveal the presence of the 2H/3R polytypism in these nanostructures. Then, the nature of their edge, surface, and bulk plasmonic excitations is fingerprinted by means of spatially-resolved electron energy-loss spectroscopy (EELS).[17,18] We trace back the origin of relevant features in the EELS spectra to specific structural characteristics of the nanoflowers, and in particular we identify contributions associated to the surface and interlayer couplings as well as to the edges of the WS$_2$ petals. The combination of EELS measurements with machine learning is then exploited to chart the ultra-low-loss region ($\Delta E \leq 5$ eV) and determine that the 2H/3R polytype of bulk WS$_2$ exhibits a semiconductor behavior with an indirect band gap of value $E_{BG} = 1.6^{+0.3}_{-0.2}$ eV.

Further, EELS spectra recorded in a monochromated TEM[19] are used to demonstrate the presence of energy-gain peaks in the WS$_2$ nanoflowers. These features could be associated with the interactions of collective excitations in the material, such as plasmons, or excitons, with the fast electrons from the beam. Remarkably, for specific nanostructures these energy-gain peaks can become more intense than their energy-loss counterparts, revealing the presence of an underlying mechanism whereby the beam electrons are on average accelerated after crossing the specimen. Such remarkable feature could be exploited to develop novel cooling strategies for atomically thin TMD nanostructures and for the devices built upon them.

## RESULTS and DISCUSSION

### *Synthesis and morphology.*

**Fig. 1a** displays a schematic diagram summarizing the synthesis of the WS$_2$ nanoflowers. These nanostructures were directly grown onto a microchip made of silicon as a frame with a silicon nitride (Si$_3$N$_4$) window in the middle (see **Fig. 1b**)[20,21]. First, tungsten trioxide (WO$_3$) powder (50 mg) was dispersed in 1mL of isopropanol (ISO). Second, a few droplets of the mixed WO$_3$ and ISO were deposited onto the microchip using a pipette (**Fig. 1c**). As indicated in **Fig. 1a**, a crucible holding 400 mg of sulfur was placed upstream from the microchip in the low-temperature region of the three-zone furnace. An argon gas flow was used both to prevent oxidation as well as to transport the sulphur vapor to the microchip. Further details of the growth procedure can be found in the Supplementary Information (SI).

**Fig. 1d** displays low-magnification annular dark-field (ADF) scanning transmission electron microscopy (STEM) images of the as-grown WS$_2$ nanostructures. We observe that the nanostructures exhibit flower-like morphologies composed by randomly oriented flakes (the "petals") arising from a common point (the "stem"), see **Fig. 1e**. The chemical composition of the WS$_2$ nanoflowers was confirmed by means of energy-dispersive X-ray spectroscopy (Supplementary **Fig. 3**).

### *Mixed 2H/3R polytypes in WS$_2$ nanoflowers.*

**Figs. 2a-b** display respectively low-magnification ADF- and bright-field (BF-) STEM images of the tip region of one representative WS$_2$ petal. The variations in image contrast indicates the presence of terraces with different number of layers each, and thus of different thicknesses ranging from 2 nm up to 30 nm. **Fig. 2c** then displays the atomic-resolution ADF-STEM image corresponding to the same petal. The petal has been oriented along the [0001] direction in order to ascertain the underlying crystalline structure. Each bright spot corresponds to an atomic column that is composed of alternating tungsten (W) and sulfur (S) atoms.

Based on the observed atomic arrangement, a hexagonal honeycomb with an atom in its center, the atomic-resolution image appears to suggest a 3R phase (Supplementary **Fig. 4b**). However, The ADF linescan extracted from the atomic resolution image across six lattice points (**Fig. 2d**) indicates a three-fold periodicity which is inconsistent with the 3R phase. Such periodicity suggests instead a layer stacking order of the type BAA', characteristic of a mixture of the 2H (AA') and 3R (AB) phases[22]. As displayed in **Fig. 2e**, this stacking sequence exhibits a broken inversion symmetry. Such polytypism has already been reported in related material like $MoS_2$[12]. As mentioned above, the 2H and 3R phases in their bilayer form are characterized by different stacking sequences, namely AA' and AB (Supplementary **Fig. 4**). In the 2H phase, the W atom is aligned with the S one and the two layers exhibit inversion symmetry. In the 3R phase instead, the W atom is staggered over S, resulting in a stacking sequence lacking inversion symmetry. Therefore, a mixture of the two phases follows the BAA' sequence, where the third layer is staggered with the S atoms over the W atoms of the second layer.

*Fingerprinting electronic excitations with spatially-resolved EELS.*

To investigate the nature of collective electronic excitations such as surface and edge plasmons arising in the 2H/3R polytypism spatially-resolved EELS is used. **Figs. 3a,b** display BF- and ADF-STEM images of horizontal $WS_2$ flakes (the nanoflower petals) overlapping among them. Some of these petals are composed by terraces of different thicknesses, as indicated by the contrast variations in the ADF-STEM image, **Fig. 3b**.

Since these petals are suspended on vacuum, it is possible to characterize their edge, surface, and bulk plasmons while reducing the contribution from coupling effects related to the nanostructure support film. We emphasize that in this geometric configuration the incident electron beam direction is close to the *c*-axis of the WS$_2$ petals.

**Fig. 3c** displays the spatially-resolved EELS map recorded in the region marked with a grey rectangle in **Fig. 3b**, where each pixel corresponds to the intensity of an individual EEL spectrum. **Figs. 3d-g** display the intensity maps associated to the same EEL signals now integrated in specific energy-loss windows: [2.5, 4.5] eV, [8, 9] eV, [12.5, 20] eV, and [21.5, 25] eV respectively. In these maps, the brighter regions highlight the dominant features appearing for each specific energy-loss range.

To better isolate the main features displayed by **Figs. 3d-g**, representative EEL spectra associated to locations along the horizontal petals indicated in **Fig. 3b** are presented in **Figs. 3h,i**. A common feature of the two sets of spectra is the appearance of a prominent peak around 23 eV that can be associated with the WS$_2$ bulk plasmon[23]. The intensity of this peak decreases as one moves from thicker towards thinner regions, following the directions of the arrows in **Fig. 3b**. This effect is particularly marked in **Fig. 3i,** corresponding to a petal composed by terraces of different thicknesses.

From the intensity map corresponding to energy losses integrated between 12.5 and 20 eV, **Fig. 3f**, one observes a significant contribution arising from the regions constituted by two different overlapping petals. The individual spectra display indeed a shoulder-like feature located around 17 eV, which becomes more marked in the thinner regions. The free electron gas dielectric response theory predicts that the surface plasmon energy $E_s$ should relate to that of the bulk one, $E_p$, by $E_s = E_p/\sqrt{2}$. In our case, this prediction corresponds to around $E_S = 16\,\text{eV}$, consistent with the experimentally observed value and confirming the surface plasmon nature of this shoulder-like feature[9].

The individual spectra in **Fig. 3h,i** also display less intense peaks located at energy losses around 3 and 8 eV. From the integrated EELS maps of **Fig. 3d** one observes that the 3 eV peak can be mainly associated to the edges of the $WS_2$ petals. Indeed, the largest signals in this energy-loss window arise from the tip of the edges and the adjacent regions. Further, we observe that the peak at 8 eV receives its main contribution associated to the interlayer coupling between petals (**Fig. 3e**). Indeed, this peak could be associated to the six $\pi$ electrons (four of which from the sulfur atoms) which are responsible for the interlayer interactions[24,25].

**Figs. 4a,b** display the ADF-STEM image and the corresponding spatially-resolved EELS map of another representative $WS_2$ nanoflower composed now by both flat and tilted $WS_2$ petals. In the latter case, their *c*-axis is close to being perpendicular to the direction of the electron beam. **Figs. 4c,d** compare the energy-loss functions for the locations indicated in **Fig. 4b** and recorded along a series of flat and tilted $WS_2$ petals respectively. We note that spectra 6, 9, and 10 in **Fig. 4b** are recorded in vacuum but close enough to the petal (aloof configuration) to identify possible contributions from its surface plasmons. We observe that the 3 eV peak appears both in the flat and tilted petals, as well as in the vacuum location sp9. In the vacuum region close to the sample, a peak for energy losses of around 3 eV is observed, whose location shifts rightwards to 3.5 eV when moving from the non-penetrating (vacuum) to the penetrating ($WS_2$ petal) locations.

Concerning the higher-energy peaks, one finds that the tilted petal configuration induces a strong coupling between the surfaces which results into a split of the 17 eV surface mode into a peak at around 20 eV and a forest of peaks between 5 and 15 eV. In **Fig. 4d,** a two-gaussian model fit has been performed to sp1 (flat) and sp8 (tilted) in the range between 10 and 35 eV. The two resulting gaussian distributions can be associated to the contributions from the bulk and surface plasmons and show how in the latter case the surface contribution at 17 eV blueshifts towards around 20 eV while becoming more marked as compared to the bulk one. We point out that related phenomena have been reported for $WS_2$ nanotubes with very small radii[23,26].

***Band gap determination of polytypic 2H/3R WS$_2$ from machine learning.***

The profile of the low-loss region of EEL spectra contains valuable information concerning the magnitude and type of the band gap. Exploiting this information requires the subtraction of the zero-loss peak (ZLP) contribution. Here the band gap of polytypic 2H/3R WS$_2$ nanoflowers is investigated by means of a recently developed model-independent method based on machine learning[27]. This approach, developed in the context of high energy physics[28], combines artificial neural networks for an unbiased parametrization of the zero-loss peak profile with Monte Carlo sampling for a faithful uncertainty estimate.

The original and subtracted EEL spectra corresponding to a representative location of a petal (labelled 5) and indicated in Supplementary **Fig. 5**, are displayed in **Fig. 5a** together with the calculated ZLP. This location corresponds to a relatively thick region where bulk behavior is expected. The uncertainty bands indicate the 68% confidence level intervals associated to the ZLP prediction and the subtracted spectrum. A functional form of the type $I_{\text{EEL,sub}}(\Delta E) = A\,(\Delta E - E_{BG})^b$ is then fitted to the subtracted spectrum ensuring that all relevant sources of uncertainty are considered[29]. The results of such fit are displayed in the inset of **Fig. 5a**.

**Fig. 5b** displays the ratio of the EELS intensity derivative, $dI_{ws2}/d(\Delta E)/\ dI_{vac}/d(\Delta E)$ between the specimen spectrum and the corresponding vacuum average. This ratio highlights how for energy losses of 1.8 eV the two profiles start to differ, indicating the onset of the inelastic contributions to the spectra. The best-fit model values are $E_{BG} = 1.6^{+0.3}_{-0.2}$ eV for the band gap value and $b = 1.3^{+0.3}_{-0.7}$ for the exponent. The latter result favors an indirect band gap, for which $b \approx 3/2$ is theoretically expected. The obtained value and nature of the band gap of bulk WS$_2$ is consistent with previous determinations in the literature.[8,14] Statistically comparable results for the model fits are obtained for other locations in Supplementary **Fig. 5**.

*Energy-gain peaks in EEL spectra.*

Additional high-resolution EELS measurements in these nanostructures were carried out using a monochromated electron microscope (see Methods). **Fig. 6a** displays the ZLP profile acquired in the vacuum; its full width at half maximum (FWHM), indicating the spectral resolution under these operational conditions, is 57 meV. The resulting EELS spectra recorded at three representative petals of the WS$_2$ nanoflowers are shown in **Figs. 6b,c,d.** Each of these spectra corresponds to petals (displayed in the insets) characterized by different shape-morphology configurations: a flat petal tip (**b**); a slightly bended petal (**c**); and a completely bended petal (**d**) respectively, where the darker contrast indicates bended regions.

These spectra exhibit both energy-gain ($\Delta E < 0$) and energy-loss ($\Delta E > 0$) peaks, distributed symmetrically about $\Delta E = 0$. The spectra in **Figs. 6b,c** correspond to WS$_2$ petals oriented randomly, with the former being flat and the latter somewhat bended. From **Fig. 6b** one finds symmetric peaks at $\Delta E = \pm 0.4$ and $\pm 0.6$ eV and a shoulder-like structure at $\Delta E = \pm 0.2$ eV. A similar pattern is observed in the thicker, slightly bended specimen of **Fig. 6c,** where again strong symmetric peaks are observed at $E = \pm 0.4$ eV but now the energy interval between the low-loss region peaks is around $0.4\ eV$. The spectrum displayed in **Fig. 6d** corresponds instead to a completely bended petal. For this configuration, the symmetric peaks appear for $\Delta E = \pm 0.25, \pm 0.35$, and $\pm 0.6$ eV. Since the positions of these peaks does not correspond to the phonon resonances of $WS_2$ [30], their origin could be associated to the interactions between the fast electron beam with the evanescent field that is created through the collective excitations of the $WS_2$ nanostructures[31]. In this respect, work in progress based on *ab initio* calculations will shed more light on the underlying mechanism responsible for these gain/loss peaks.

The ratio between the gain and loss peak intensities, $I_G/I_L$, is known to depend on the relation between the associated thermal energies[32]. From the spectra displayed in **Figs. 6b,c,d,** one observes that this ratio is typically larger than one. Specifically, for the symmetric peaks closer to the ZLP in each spectrum, values of $I_G/I_L$ = 1.2, 1.6, and 1.5 are obtained respectively. We find that bended petals appear to exhibit larger gain/loss ratios than their flat counterparts. We note that related features have been also observed in TMDs when studied by means of Raman scattering[33], where anomalously intense anti-Stokes peaks with $I_G/I_L > 1$ are reported. Such remarkable behavior could be exploited to develop novel cooling strategies for atomically thin TMD nanostructures and devices built upon them.

**CONCLUSIONS**

Our understanding of the consequences of unconventional mixed crystalline phases on the properties of TMD nanomaterials remains partially *terra incognita*. In this work, we have carried out an exhaustive characterization of WS$_2$ nanoflowers which represent an ideal laboratory for the study of the modifications of local electronic properties of WS$_2$, thanks to their mixed 2H/3R polytypism and rich variety of shape-morphology configurations. By means of state-of-the-art AC-STEM and EELS measurements, we have fingerprinted the nature of their edge, surface, and bulk plasmonic excitations and traced back the origin of distinctive features in the EELS spectra to specific structural characteristics of the nanoflowers.

Further, thanks to a recently developed ZLP subtraction method based on machine learning, we have stablished that the 2H/3R polytype of bulk WS$_2$ exhibits a semiconductor behavior with an indirect band gap of $E_{BG} = 1.6^{+0.3}_{-0.2}$ eV. We have also demonstrated the presence of energy-gain peaks in EELS spectra recorded in WS$_2$ nanoflowers and characterized by a gain-to-loss ratio $I_G/I_L > 1$, indicating an underlying mechanism whereby electrons are on average accelerated after crossing the specimen. Such feature could be exploited to develop novel cooling strategies for atomically thin TMD nanostructures and devices built upon them. Our results represent a stepping stone in a program aiming to develop a systematic strategy for the controllable growth of TMD nanostructures characterized by mixed crystalline phases.

**Fig. 1: Schematic of the synthesis and morphology of WS₂ nanoflowers.**

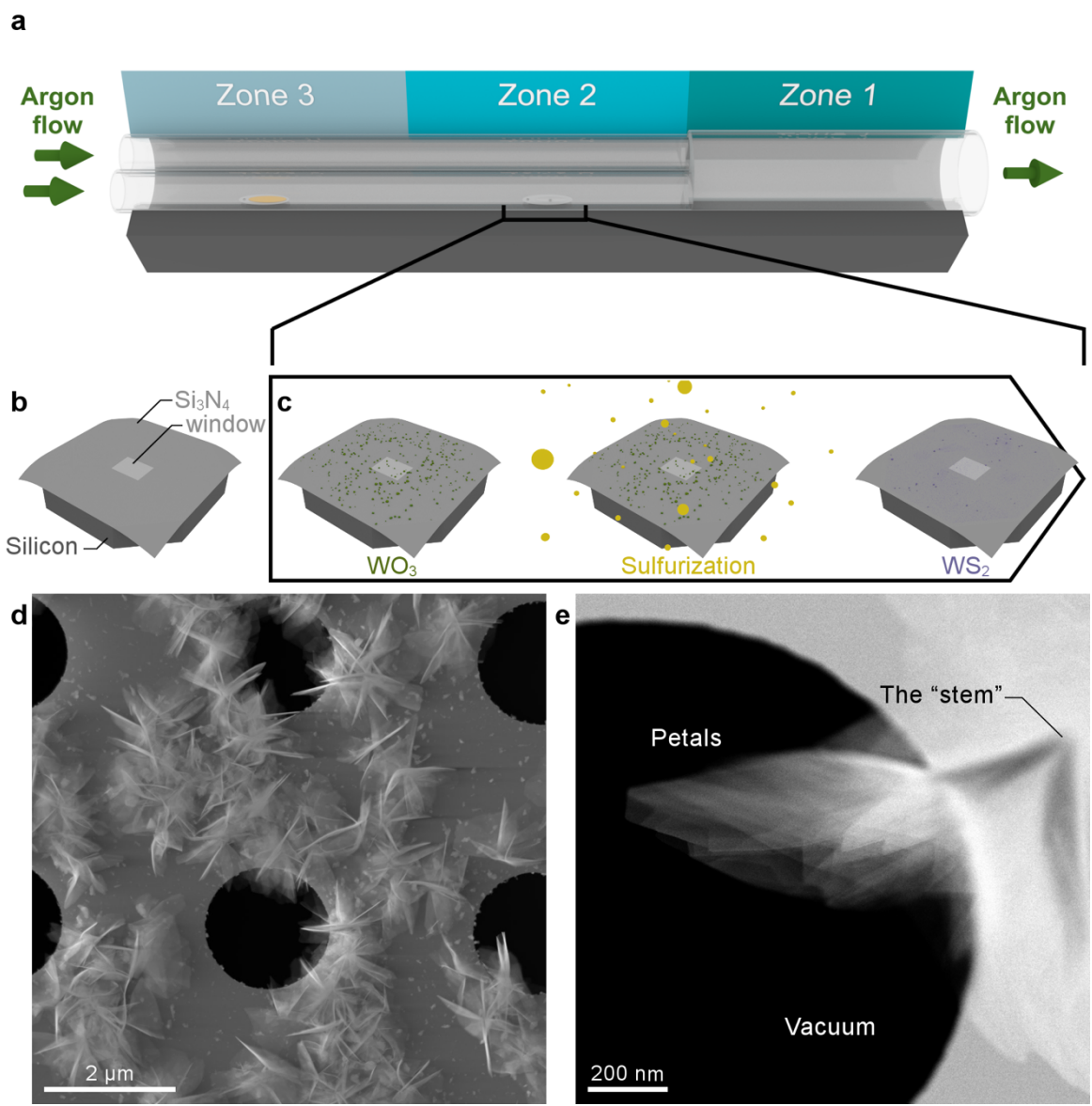

**a** – **c**, Schematic illustration of the synthesis procedure for the WS$_2$ nanoflowers within the three-zone furnace. **d**, Low-magnification ADF-STEM top-view perspective of the as-grown WS$_2$ nanostructures illustrating their flower-like morphology. The nanoflowers are grown on top of a holey Si$_3$N$_4$ TEM grid. **e**, ADF-STEM image of one representative WS$_2$ nanoflower composed by free-standing, randomly oriented WS$_2$ flakes (petals) protruding from the stem of the flower. This configuration is especially advantageous to eliminate the contributions related to the nanostructure-support film coupling effects.

**Figure 2: AC-STEM Z contrast images of the mixed 2H/3R polytypes in WS$_2$ petals.**

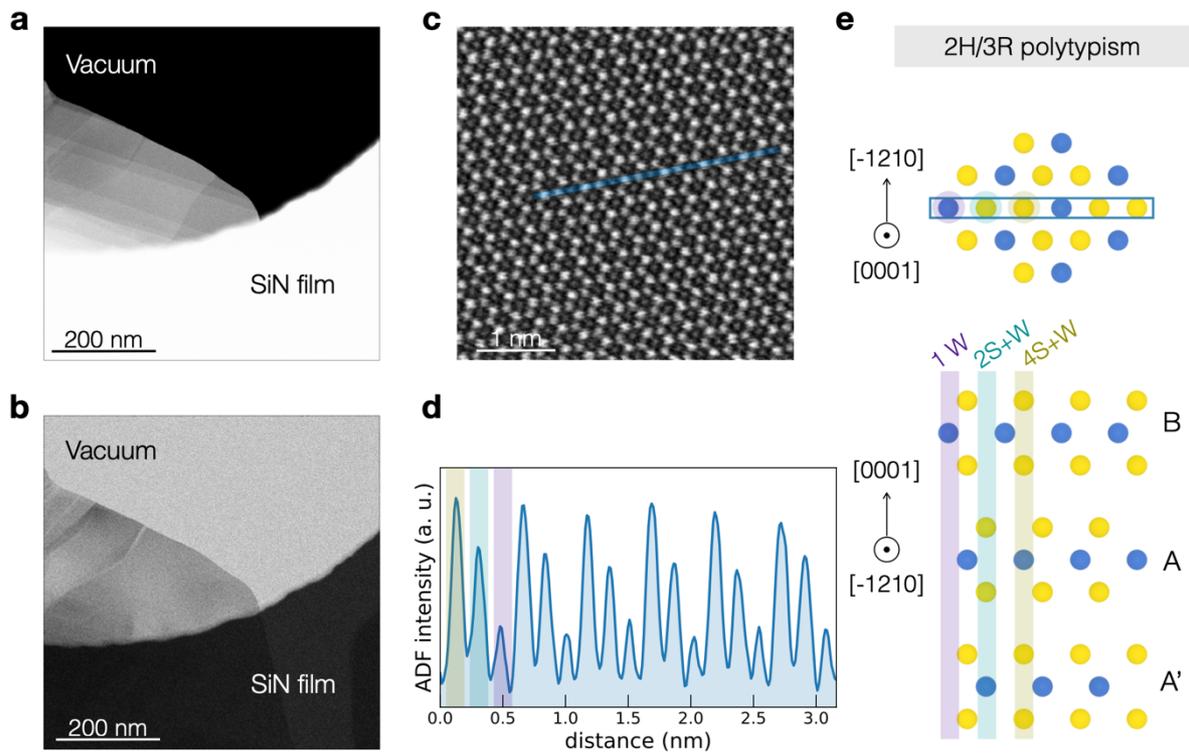

**a,b** Low-magnification ADF- and BF-STEM images respectively of a representative WS$_2$ petal, where we also indicate the vacuum and SiN film regions. The difference in contrast observed in the petal of **a** corresponds to terraces of different thicknesses. **c**, Atomic resolution image corresponding to the petal tip area. **d**, The ADF intensity profile acquired along the blue region in **c**. **e**, Schematic atomic model of the top-view (upper panel) and side-view (lower) of the crystalline structure associated to the mixed 2H/3R polytypism.

**Fig. 3: Spatially-resolved EEL spectra in flat WS$_2$ petals.**

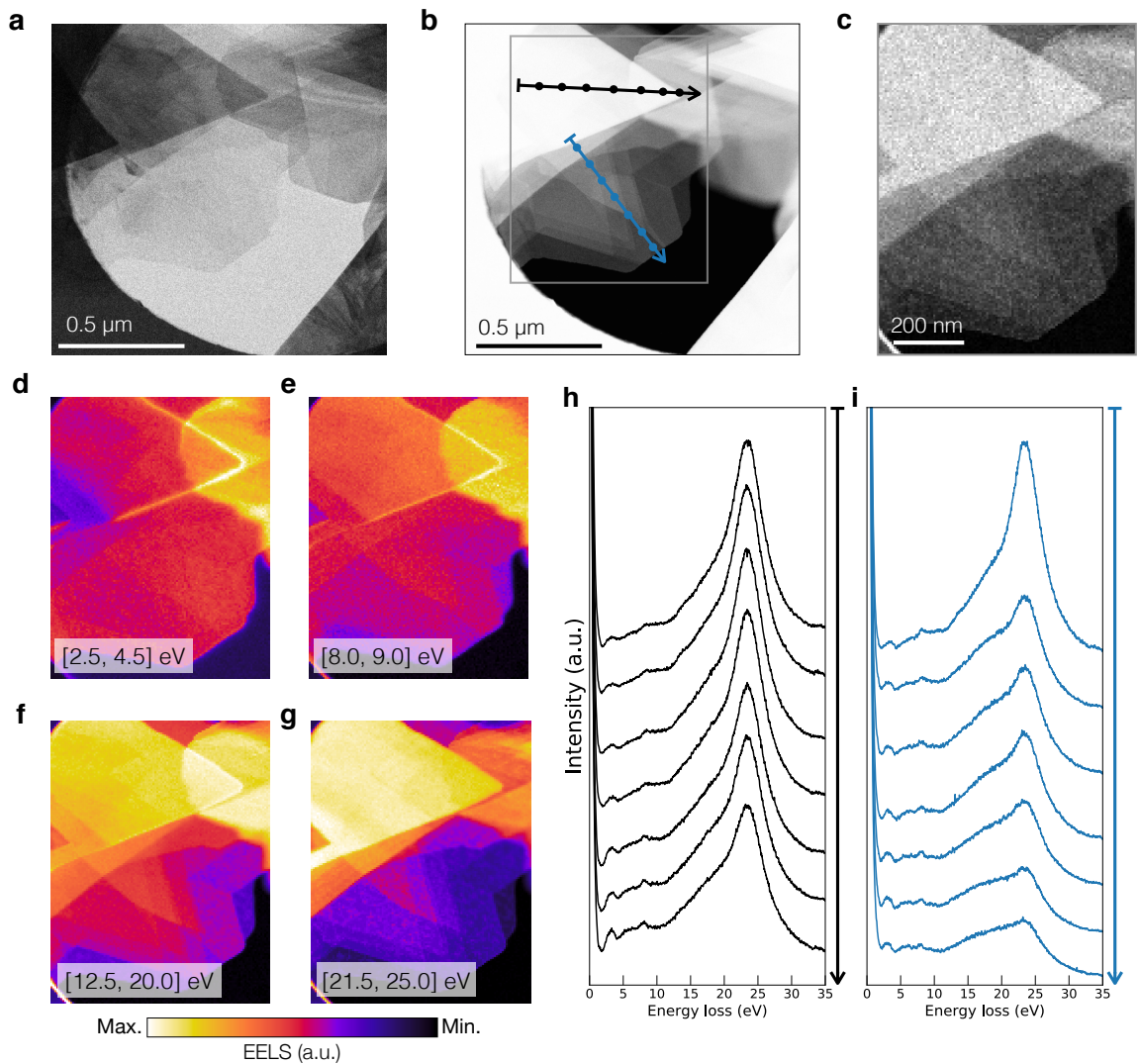

**a,b**, BF- and ADF-STEM images respectively of horizontal (flat) WS$_2$ petals. **c,** Spatially-resolved EELS map of the area indicated with a grey rectangle in **b**. **d-g**, Intensity maps of the EELS signals integrated for different energy-loss windows. **h, i**, Individual EEL spectra taken at different locations along the horizontal petals as indicated by the black and blue arrows in **b**. The spectra displayed in **h** correspond to a region characterized by the overlap of individual petals arising both from the left and the right sides of **a,b**. The spectra in **i** corresponds instead to a region displaying a petal composed by different terraces and thus different thicknesses.

**Fig. 4: EEL spectra in tilted WS$_2$ petals.**

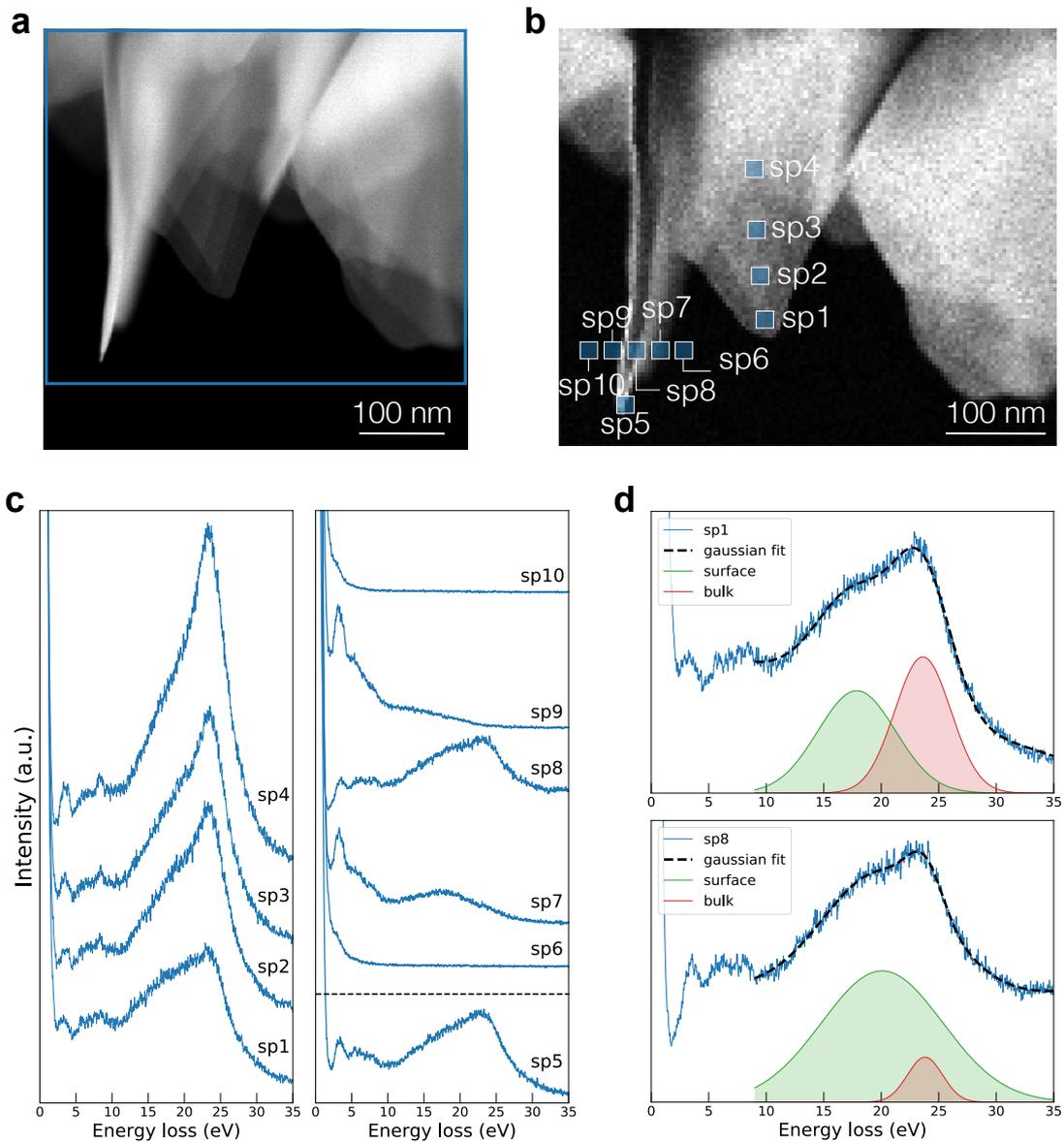

**a,** ADF-STEM image of another representative WS$_2$ nanostructure composed both by flat and tilted WS$_2$ petals. **b**, Spatially-resolved EELS map of the area indicated with a blue rectangle in **a**. **c**, Individual EELS spectra corresponding to the locations indicated in **b**. Spectra 1 to 4 are associated to a flat region of the WS$_2$ petal, while spectra 6 to 10 correspond to locations that cross the tilted WS$_2$ petal, with sp5 recorded at its tip. Note that sp6, 9, and 10 are recorded in vacuum but close enough to the petal to identify possible contributions from its surface plasmons. **d,** Spectra 1 and 8 where a two-gaussian model fit has been performed in the range between 10 and 35 eV. The two resulting gaussian distributions are also displayed and can be associated to the contributions from the bulk and surface plasmons.

**Fig. 5: Band gap determination in polytypic WS$_2$ from machine learning.**

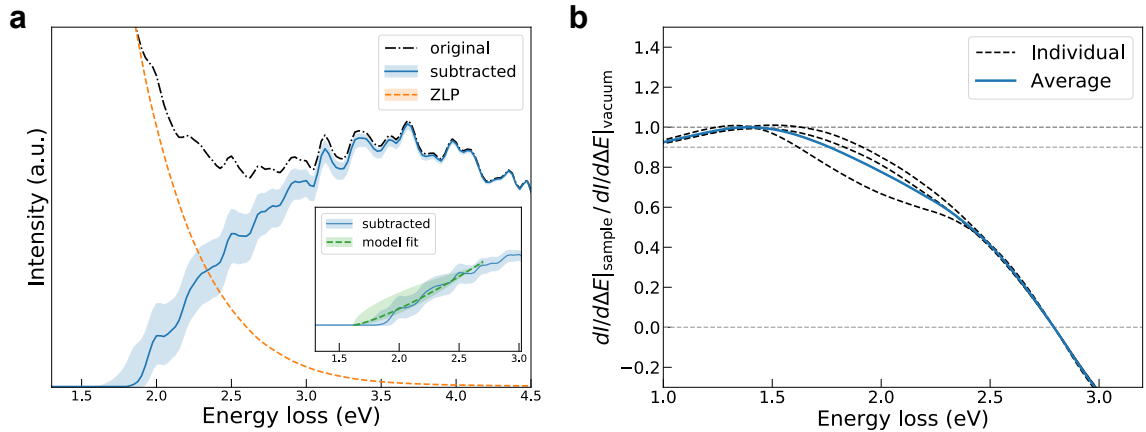

**a**, The low-loss region of the original EEL spectrum recorded at the location indicated in the Supplementary **Fig. 5** together with the resulting subtracted spectrum. We also display the ZLP model prediction used for the subtraction procedure. The inset shows the result of the polynomial fit to the onset region of the subtracted spectrum and the bands represent the 68% CL intervals for the model uncertainties. **b**, The ratio of the derivative of the original EELS intensity profile indicated in **a** over the average of the corresponding derivative for spectra recorded in the vacuum. The value of the energy loss for which this ratio differs from unity by more than 10% determines the training range of the machine learning model. Note that this ratio crosses the *x* axis at the first local minimum of the unsubtracted spectrum.

**Figure 6. Energy-gain peaks in EEL spectra.**

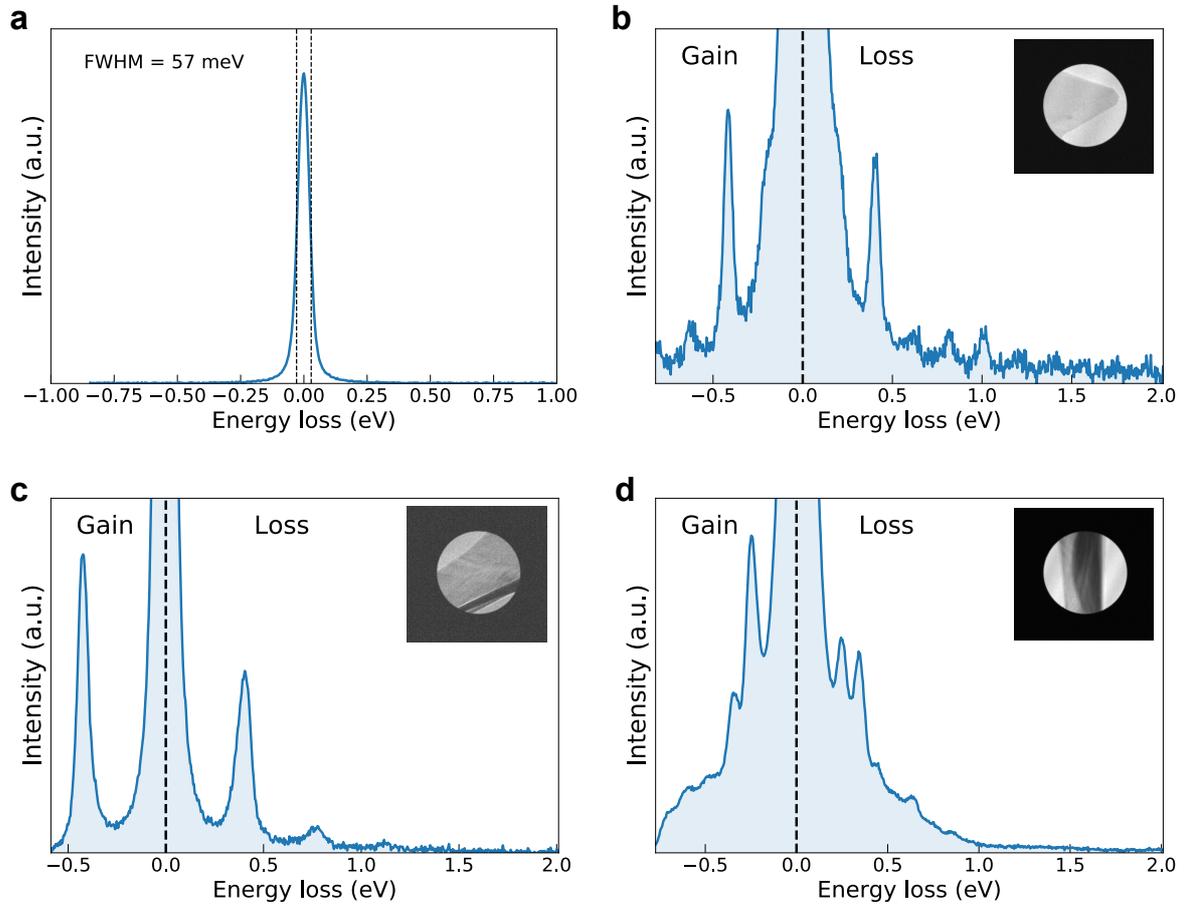

**a**, The ZLP recorded in the conventional TEM mode adopted for the vibrational spectroscopy measurements. The value of its FWHM indicates a spectral resolution (FWHM) of 57 meV. **b,c,d,** The electron energy gain/loss spectra for three representative $WS_2$ petals with different shape-morphology configurations: a flat petal tip (**b**); a slightly bended petal (**c**); and a completely bended petal (**d**) respectively. The insets display the TEM images for the area where each spectrum was recorded, where a darker contrast indicates a bended region of the petal. The gain/loss peaks appear symmetrically with respect to $\Delta E = 0$.

**METHODS**

*Electron microscopy characterization.* The low magnification ADF image in **Fig. 1d** and the EDS maps shown in Supplementary **Fig. 3** were taken using a Titan Cube microscope operated at 300 kV. The ADF- and BF-STEM and the EELS measurements displayed in **Figs. 1e, 2a-c, 3,** and **4** were taken using a JEOL ARM200F microscope with a cold field-emission gun operated at 60 kV. This microscope is equipped with an aberration probe corrector and a Gatan GIF Quantum spectrometer. The convergence and collection semi-angles were set to 30.0 mrad and 66.7 mrad, respectively. The spectral resolution (FWHM of the zero-loss peak) achieved under these conditions is around 450 meV. The vibrational spectroscopy measurements in **Fig. 6** were collected in a ARM200F Mono-JEOL microscope equipped with a GIF continuum spectrometer. The microscope was operated at 200 kV in TEM mode with the monochromator ON and the smallest slit of 0.1 um inserted. The aperture of the GIF was set to 1 mm. Under these conditions, we achieved a spectral resolution (FWHM) of 57 meV. In TEM-EELS the angular distribution of the electrons entering to the spectrometer aperture is independent of the entrance aperture size. The angular distribution is controlled by the size of the objective aperture in the diffraction pattern formed in the back focal plane of the objective lens. For these measurements, an objective aperture of 40 μm was used, therefore the collection angle becomes around 6 mrad.

**Data availability**

All data generated or analyzed in this study can be obtained from the authors upon request.

**Acknowledgements**

We are grateful to Emanuele R. Nocera and Jacob J. Ethier for assistance in installing EELSfitter in the Nikhef computing cluster. We acknowledge discussions with Javier Garcia de Abajo on energy-gain phenomena in TMDs.

**Funding**

S.E.v.H. and S.C.-B. acknowledge financial support from the ERC through the Starting Grant "TESLA", grant agreement no. 805021. L.M. acknowledges support from the Netherlands Organizational for Scientific Research (NWO) through the Nanofront program. The work of J. R. has been partially supported by NWO.

**Author information**

*Contributions*

S.E.v.H. performed the fabrication of the specimen. M.M, E.O, and H.H carried out the electron microscopy measurements. S.C.-B. and S.E.v.H. analyzed the electron microscopy data. L.I.R., L.M., and J.R. developed the machine learning approach for the subtraction of the zero-loss peak and evaluated the band gap. S.C.-B designed and supervised the experiments and coordinated the data assembly. S.C.-B and S.E.v.H. wrote the paper and edited the figures, while all the authors contributed to the discussion.

*Corresponding author*

Correspondence to Sonia Conesa-Boj (s.conesaboj@tudelft.nl)

**Ethics declarations**

*Competing interests*

The authors declare no competing interests.

Supplementary Information

# Illuminating the electronic properties of WS$_2$ polytypism with electron microscopy

Sabrya E. van Heijst[1], Masaki Mukai [2], Eiji Okunishi[2], Hiroki Hashiguchi[2], Laurien I. Roest[1,4], Louis Maduro[1], Juan Rojo[4,5], and Sonia Conesa-Boj[1,*].

[1] Kavli Institute of Nanoscience, Delft University of Technology, 2628CJ, Delft, The Netherlands.

[2] EMBU JEOL Ltd, Tokyo, Japan.

[4] Nikhef Theory Group, Science Park 105, 1098XG Amsterdam, The Netherlands.

[5] Department of Physics and Astronomy, VU Amsterdam, 1081 HV Amsterdam, The Netherlands.

*e-mail: s.conesaboj@tudelft.nl

**Table of contents**

## A. Fabrication of the WS₂ nanoflowers.

Supplementary **Fig. 1a** provides a schematic illustration of the gradient tube furnace used for the fabrication of the WS$_2$ flower-like nanostructures. This furnace consists of a long quartz reactor tube placed inside an electric oven. The latter has three separate heating elements, creating three heating zones within the furnace which enables individual temperature control. Due to limited thermal isolation, heating might cause a temperature increase in the neighboring zone(s) when the temperature gradient becomes too large. A gas system is in place which allows an argon flow from the source, connected to the entrance (at zone 3), to the exhaust at the backend (at zone 1). Using a flow meter, the argon flow can be regulated in a range from 100 to 1000 sccm. As the argon flow is split between the two tubes present in the system, each tube will receive half the argon flow set on the flow meter. At the backend of the furnace the flow is led through a bubbler system before being disposed of via the exhaust.

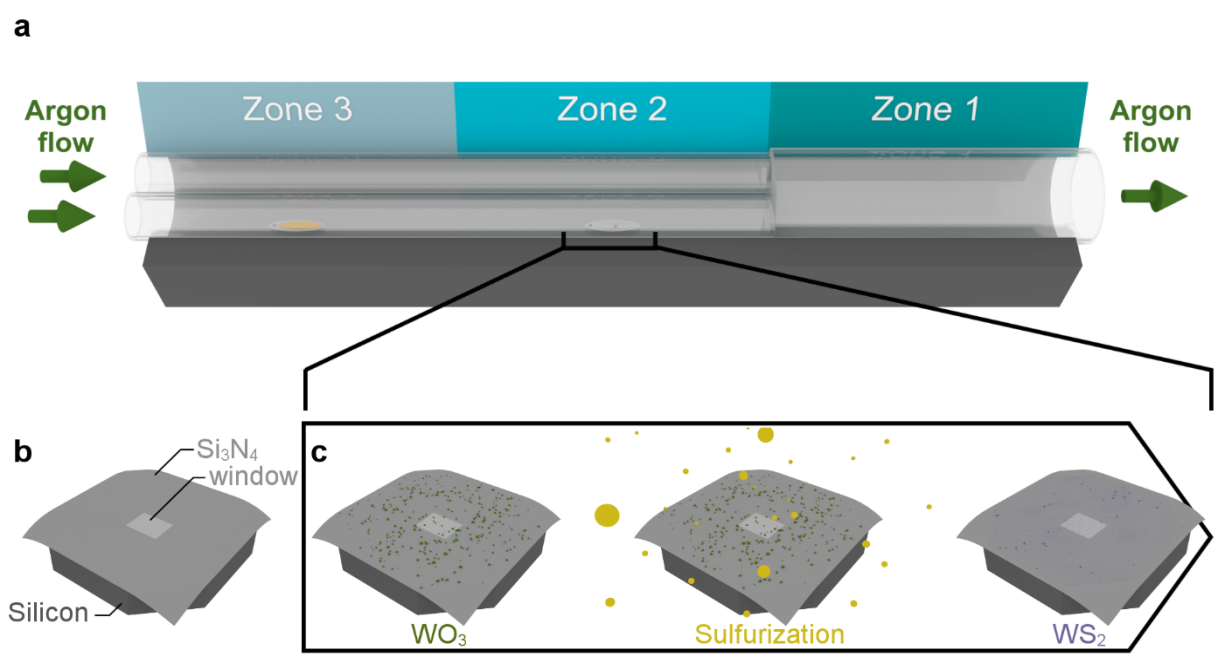

**Supplementary Fig. 1: a**, Schematic illustration of the three-zone gradient tube furnace. **b**, The microchip on which the WS$_2$ nanoflowers are grown. It is composed by a silicon frame and a silicon nitride film, which is spanned freely over a window in its center. **c**, Schematic representation of the synthesis procedure used in the fabrication of the WS$_2$ nanoflowers.

The growth protocol adopted for the fabrication procedure is schematically represented in supplementary **Fig. 2**. Preceding the sulfurization process (Supplementary **Fig. 1c**), the system undergoes a pre-treatment to rid it of oxygen remnants. Herein the system is flushed with a 500 sccm argon flow for 30 minutes. Following this pre-treatment, the temperature of zone 2 (the region in which the microchip is situated) is ramped up at a target rate of 10 °C/min. This temperature ramping is continued until zone 2 is at the desired temperature of 750 °C.

The argon flow is reduced from 500 sccm to 150 sccm once a temperature of 500 °C has been attained.

Once the target reaction temperature has been reached, zone 3, containing the sulfur precursor, is heated to a temperature of 220 °C. At a reaction temperature of 750 °C, this target temperature of 220 °C is already obtained prior to arriving at the desired reaction temperature in zone 2 due to the effects of limited thermal isolation. The initial reaction time is therefore defined as the moment for which the target reaction temperature of 750 °C is reached in zone 2. The total reaction time is thus measured from this point onwards. The furnace is held at these reaction conditions for 1 hour. After this time, the furnace is left to cool down naturally down to room temperature under continued argon flow.

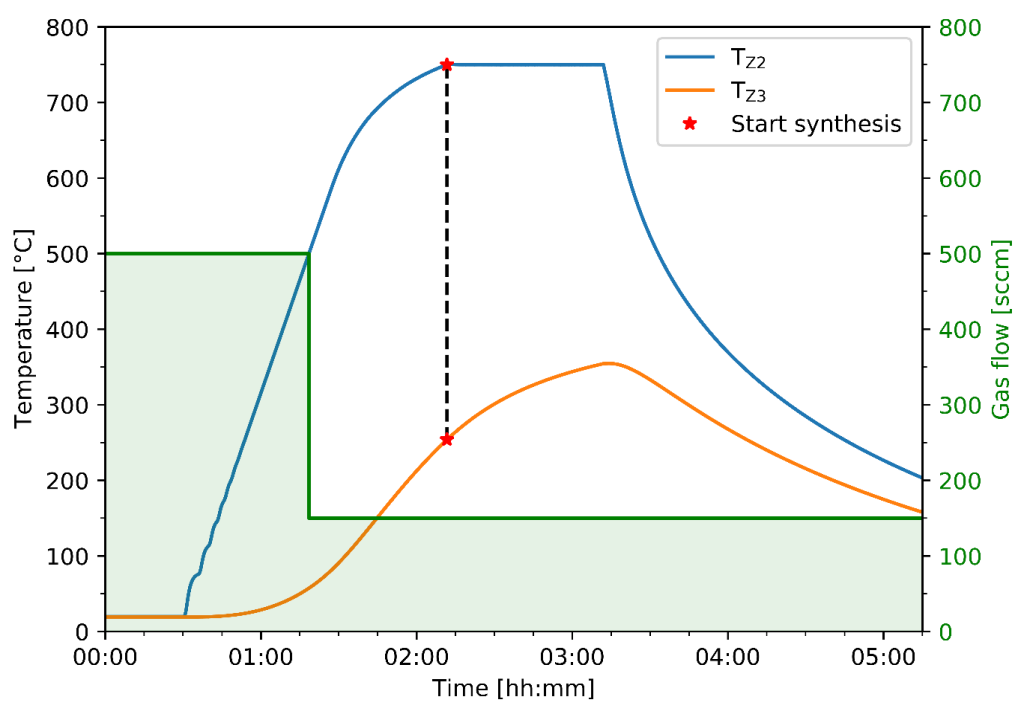

**Supplementary Fig. 2:** We display the temperature (left *y*-axis) and the argon gas flow (right *y*-axis) as a function of the reaction time. In this scheme, $T_{Z2}$ and $T_{Z3}$ indicate the temperatures of zones 2 and 3 of the furnace, respectively. The stars indicate the temperatures attained in zones 2 and 3 at the starting time of the reaction.

## B. Energy-dispersive X-ray spectroscopy (EDS) characterisation.

Energy-Dispersive X-ray Spectroscopy (EDS) measurements were carried out in order to verify of the chemical composition of the nanoflowers. Supplementary **Fig. 3a** displays a low-magnification ADF-STEM image of the area to be inspected by EDS mapping. Within this area, multiple flower-like nanostructures are visible. The corresponding EDS maps of tungsten (blue) and sulfur (yellow) are shown in Supplementry **Fig. 3b**. The EDS characterization of the sample demonstrates the presents of both tungsten and sulfur in the nanostructures thus confirming the composition to the nanoflowers to be $WS_2$.

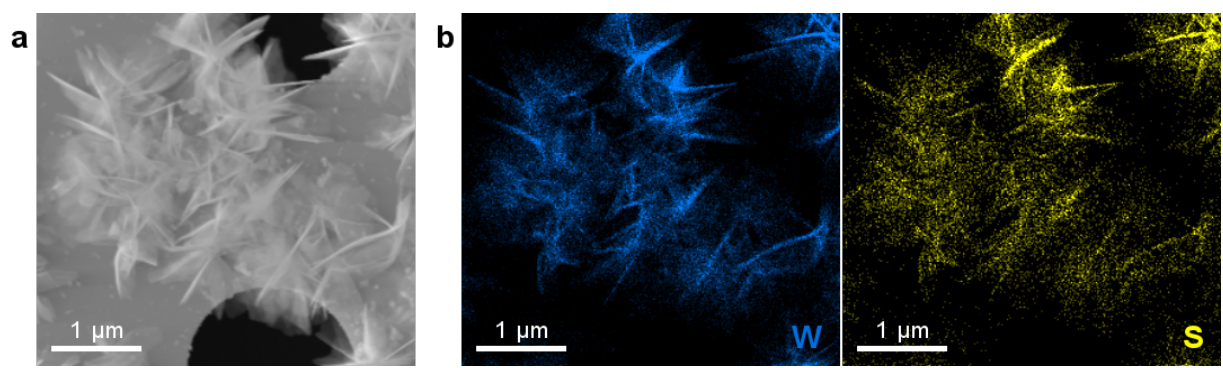

**Supplementary Fig. 3: a**, Low-magnification ADF-STEM images of $WS_2$ nanostructures. **b,c**, EDS compositional maps of the region displayed in **a**, with the tungsten (W) and sulfur (S) maps represented in blue and yellow respectively.

## C. Stacking sequences in TMD materials.

TMD materials exist in multiple polytypes, each of which is characterized by a unique stacking sequence. Two of these polytypes are 2H and 3R, using a convention where the number denotes the number of layers in the unit cell and the letter describes the crystallographic structure. In the monolayer limit, no difference can be observed between the hexagonal (2H) and rhombohedral (3R) crystal structures. The reason is that in both polytypes the monolayer is composed of a transition metal atomic layer sandwiched between two chalcogen atomic layers.

The differences between the two polytypes become apparent for bilayer stacking (Supplemenatry **Fig. 4**). For the 2H polytype, the second layer is rotated 60° along the *c*-axis relative to the first layer, and shifted so the transition metal (chalcogen) atoms of the second layer are positioned above the chalcogen (transition metal) atoms of the first layer. This stacking sequence, AA', is depicted in Supplementary **Fig. 4a** and possesses inversion symmetry. For the case of the 3R bilayer stacking, the second layer is not rotated but there is a shift in the position which places the chalcogen atoms of the second layer above the transition metal atoms of the first layer such that the resulting stacking sequence becomes AB. The 3R bilayer stacking is shown in Supplemenatry **Fig. 4b** and one observes how in this case the inversion symmetry which characterised the 2H phase is now broken.

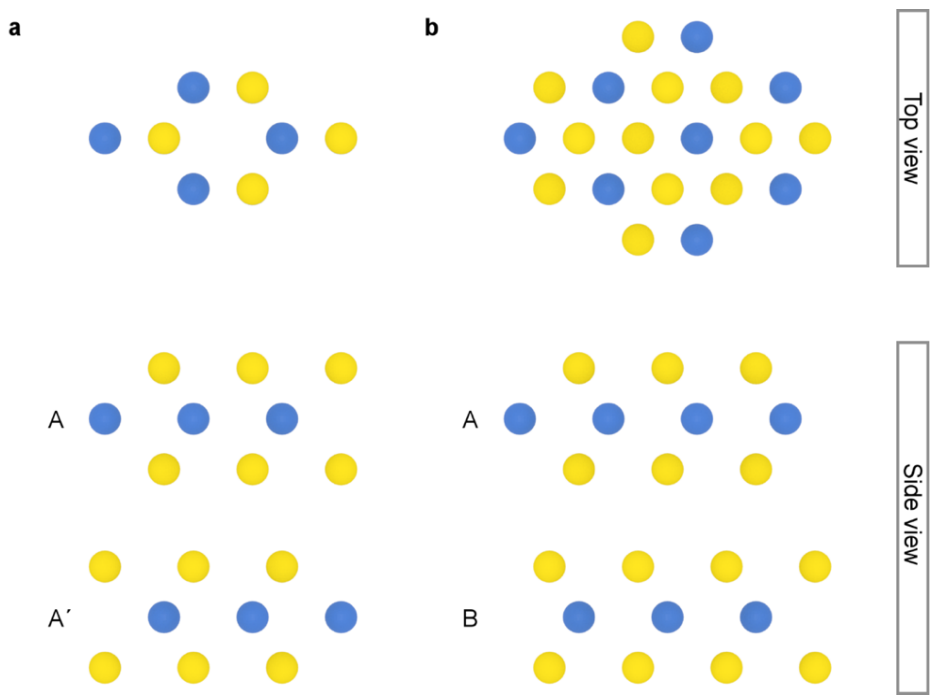

**Supplementary Fig. 4:** Schematic atomic model of the top-view (upper panels) and side-view (lower panels) of the bilayer crystalline structure associated to the 2H and 3R polytypes, **a** and, **b** respectively. Here the tungsten atoms are indicated in blue and the sulfur atoms in yellow.

## D. Band gap determination of polytypic 2H/3R WS$_2$ from machine learning.

The band gap determination of polytypic 2H/3R WS$_2$ is performed from EELS spectra acquired from the same WS$_2$ petal shown in **Fig. 2** in the main text. Supplementary **Fig. 5** displays a spatially-resolved EELS map of this petal, where locations labelled as 1, 2 and 3 (4 and 5) correspond to spectra recorded in the vacuum (the specimen). The spectrum from location 5 is used for the band gap analysis, and consistent results are obtained if the spectrum from location 4 is used instead (Supplementary **Figs. 5b** and **5c**). Further discussions on the technical procedure adoped for the band gap determination can be found in ref.[27] from the main manuscript.

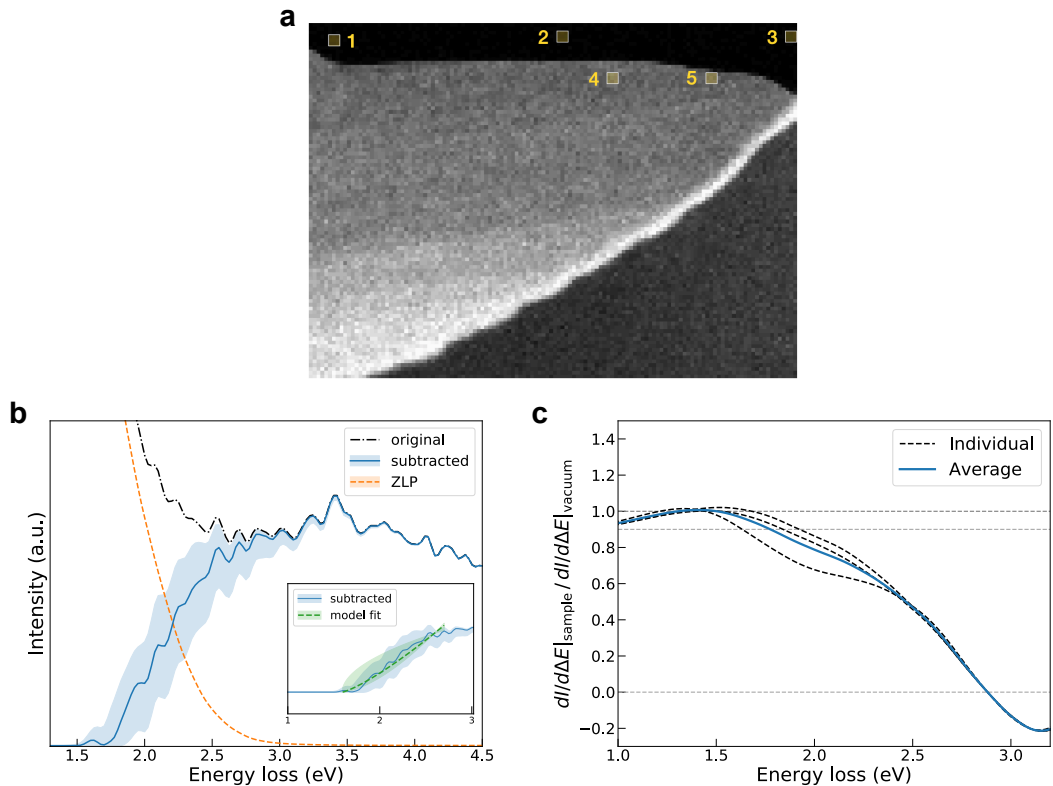

**Supplementary Fig. 5: a,** Spatially-resolved EELS map of one representative petal of a WS$_2$ nanoflower. This petal is the same one used for atomic resolution STEM image in **Fig. 2c** of the main text. The yellow squares indicate the location of the EEL spectra used to determine the band gap by means of machine learning methods. Specifically, location 5 is used in the main text, and consistent results are found with the spectrum recorded at location 4. **b,c** same as **Fig. 5** in the main manuscript now for the spectrum recorded at location 4 in **a**.